\documentclass[12pt]{article}
\usepackage{graphicx}
\usepackage{mathtools}
\usepackage{caption}
\usepackage{subcaption}

\newcommand\pubnumber{NuPhys2016-Branca}
\newcommand\pubdate{\today}

\def\padova{INFN - Sezione di Padova\\
Via Marzolo 8, 35131 Padova - Italy}
\def\email{\footnote{\emph{email:} branca@pd.infn.it}}
          
\def\Title#1{\begin{center} {\Large #1 } \end{center}}
\def\Author#1{\begin{center}{ \sc #1} \end{center}}
\def\Address#1{\begin{center}{ \it #1} \end{center}}

\newcommand\pubblock{\rightline{\begin{tabular}{l} \pubnumber\\
         \pubdate  \end{tabular}}}
\newenvironment{Abstract}{\begin{quotation}  }{\end{quotation}}
\newenvironment{Presented}{\begin{quotation} \begin{center} 
             PRESENTED AT\end{center}\bigskip 
      \begin{center}\begin{large}}{\end{large}\end{center} \end{quotation}}

\def\beq{\begin{equation}}
\def\eeq#1{\label{#1}\end{equation}}
\def\eeqn{\end{equation}}

\def\beqa{\begin{eqnarray}}
\def\eeqa#1{\label{#1}\end{eqnarray}}
\def\eeqan{\end{eqnarray}}

\let\bar=\overbar

\def\Dslash{\not{\hbox{\kern-4pt $D$}}}
\def\dslash{\not{\hbox{\kern-2pt $\del$}}}

\def\msb{{\bar{\ssstyle M \kern -1pt S}}}

\begin{document}
\begin{titlepage}
\pubblock

\vfill
\Title{The CUORE experiment at the LNGS}
\vfill
\Author{Antonio Branca\email  for the CUORE Collaboration}
\Address{\padova}
\vfill
\begin{Abstract}
CUORE is the first ton-scale experiment based on the bolometric technique to search for neutrinoless double-beta decay. Its core is made of 988 $\mathrm{TeO_{2}}$ crystals cooled down to $10~\mathrm{mK}$. The temperature must be as stable as possible during detector operations, for an integrated live-time of about 5 years. To reach these goals, a dedicated cryogenic system has been developed. Temperature stabilization of the crystals response will be performed to compensate for residual instabilities. External vibrations that can deteriorate the crystals energy resolution, are dumped thanks to a suspension system. Moreover, a thorough selection and cleaning process performed on the construction material will allow the abatement of the radioactive backgrounds. Here the key aspects of the systems are presented.
\end{Abstract}
\vfill
\begin{Presented}
NuPhys2016, Prospects in Neutrino Physics\\
Barbican Centre, London, UK,  December 12-14, 2016
\end{Presented}
\vfill
\end{titlepage}
\def\thefootnote{\fnsymbol{footnote}}
\setcounter{footnote}{0}

\section{Introduction}

If the neutrino is a Majorana particle, some atomic nuclei can undergo neutrinoless double beta decay ($\mathrm{0 \nu \beta \beta}$) \cite{0dbd}, with a straightforward signature: a mono-energetic peak, given by the sum of the energies of the two electrons in the final state, around the Q-value of the atomic transition. Nevertheless, the predicted half life $\mathrm{T^{0\nu}_{1/2}}$ of the process, greater than $\mathrm{10^{25} - 10^{26}~yr}$, makes the experimental search extremely challenging. Such a challenge is worth the effort, since a measure of $0 \nu \beta \beta$  would give information on the neutrino mass hierarchy, but would also represent a discovery of physics beyond the Standard Model. In fact, the Majorana nature of the neutrino and the lepton number violation would be proven; the latter representing a possible source of the matter-antimatter asymmetry in the early universe. 

In this proceeding, the key aspects of the CUORE experiment, that will search for $\mathrm{0 \nu \beta \beta}$ decay of the $\mathrm{\prescript{130}{}{\mathrm{Te}}}$ isotope, are discussed. 

\section{CUORE}
The CUORE experiment \cite{qexp}  (scheme shown in fig. \ref{fig:sub1}) is located at the LNGS underground facility, at an average depth of $\sim3600~\mathrm{m}$ water equivalent, to suppress the natural environmental radioactivity background (at LNGS the muon, gamma and neutron fluxes are $\Phi_{\mu} = 3\cdot 10^{-8}~\mathrm{\mu/(cm^{2} \cdot s})$, $\Phi_{\gamma} = 0.73~\mathrm{\gamma/(cm^{2} \cdot s})$, $\Phi_{n} < 4\cdot 10^{-6}~\mathrm{n/(cm^{2} \cdot s})$ below 10 MeV, respectively). It exploits $\mathrm{TeO_2}$ crystals both as source for the $\mathrm{0 \nu \beta \beta}$ decay and as particle detectors for the pair of electrons in the final state. The energy $E$ deposited by the electrons is measured through the temperature rise $\Delta T$ of the crystals. At low temperatures $\Delta T \propto (E/T^{3}) e^{-t/\tau}$, with a decay time $\tau \propto T^{3}$. The experiment is a ton-scale detector, with a core of $988$ $\mathrm{TeO_2}$ crystals with a total mass of $742~\mathrm{kg}$ arranged in $19$ towers. A massive detector is mandatory for a good sensitivity to the $\mathrm{0 \nu \beta \beta}$ process, which is $[T^{0\nu}_{1/2}]^{sens.} \propto \sqrt{M \cdot t / \Delta E \cdot B}$, where $M$ is the mass of the $\prescript{130}{}{\mathrm{Te}}$ isotope, $t$ is the exposure time, $\Delta E$ the energy resolution and $B$ the number of background events.

Taking into account the considerations above, the following main challenging experimental goals can be identified: to achieve measurable, constant and fast signals at a given energy, the $\sim 1$ ton-scale detector has to be cooled down to cryogenic and stable temperatures; a good sensitivity can be obtained only if  the detector is run in a low background environment, allowing high energy resolution and low number of background events. 

An important test-bench for the development of the CUORE experiment has been the CUORE-0 prototype \cite{q0exp}. CUORE-0 detector was a CUORE-style tower of $\mathrm{TeO_2}$ crystals, built following all the protocols developed to produce, clean and assemble the CUORE detector components, thus representing an important opportunity to test the CUORE assembly line and the material cleaning techniques. Moreover, the prototype allowed to validate both the background model and energy resolution, and to obtain physics results \cite{q0res}: the upper limit on the effective Majorana mass is shown in fig. \ref{fig:sub2} (result combined with that from the previous prototype Cuoricino). CUORE-0 obtained a total background at the region of interest (ROI) of $0.058 \pm 0.004~\mathrm{counts/(keV \cdot kg \cdot yr)}$ and an energy resolution of $5~\mathrm{keV}$ FWHM at $2615~\mathrm{keV}$.


\section{Key factors to reach the CUORE goals}

The main key factors to reach the goals of the CUORE experiment are: the cryogenic system, the suspension system and the material cleaning and assembling.

\subsection{Cryogenic system}

The CUORE cryostat consists of 6 nested vessels corresponding to different temperature stages (see fig. \ref{fig:sub1}). The outermost shield, Outer Vacuum Chamber (OVC), at room temperature, and the Inner Vacuum Chamber (IVC), at $4~\mathrm{K}$, are vacuum-tight, and are separated by an intermediate radiation shield maintained at a temperature of $\sim40~\mathrm{K}$. A system of $5$ Pulse Tubes (PTs), mounted on the OVC top plate, allow the cooling of the $40~\mathrm{K}$ shield and of the IVC. Since PTs are cryocoolers, the cryostat is cryogen-free and a high duty cycle is achievable, minimising data-taking dead times. Nevertheless, the use of only PTs would require months to bring the detector from room temperature to the operations temperature, given the detector mass. Thus a Fast Cooling System (FCS) has been developed. Helium gas is circulated through a separate cryocooler, where a system of heat exchangers cools the gas down to less than $40~\mathrm{K}$. The He gas is then injected into the IVC. This system allows to reduce the cool down period of the whole detector to few weeks. Inside the IVC there are three radiation shields at $600~\mathrm{mK}$, $50~\mathrm{mK}$ and the coldest one at $10~\mathrm{mK}$. The coldest temperature is achieved through a $\prescript{3}{}{\mathrm{He}}$/$\prescript{4}{}{\mathrm{He}}$ dilution refrigerator (DU), specifically designed for CUORE by Leiden Cryogenics. A system of temperature stabilisation has been developed to correct possible drifts of the operation temperature of $10~\mathrm{mK}$. To attenuate neutron and $\gamma$-ray backgrounds, the cryostat is surrounded by layers of borated polyethylene, boric-acid powder, and lead bricks. A further suppression of $\gamma$-rays from the cryostat materials is obtained with additional lead layers inside the cryostat, including ancient Roman lead \cite{romanpb}.

\subsection{Suspension system}

Mechanical noise is a source of background that generates energy dissipation into the crystals, worsening their energy resolution. A suspension system provides a mechanical decoupling of the detector from the outside environment. The system minimizes the transmission of mechanical vibrations due to seismic noise and operations of the cryocoolers and pumps. The detector is hung by the Y-Beam through cables made out of stainless steal tie bars, Kevlar ropes and copper bars; damping the horizontal oscillations. This part of the system has three important characteristics: is able to sustain the detector weight of $\sim 1.2~\mathrm{ton}$; has a low thermal conductivity, being in contact with the $10~\mathrm{mK}$ shield; has a low radioactive contamination, being close to the crystals. Three minus-K springs connect the Y-beam to the Main Support Plate (MSP), attenuating the noise of $\sim 35~\mathrm{dB}$.  Minus-K springs provide vertical-motion isolation by a stiff spring that supports the weight load, the Y-beam and the detector, combined with a negative-stiffness mechanism. The net vertical stiffness is made very low without affecting the static load-supporting capability of the spring. Beam-columns connected in series with the vertical-motion isolator provide horizontal-motion isolation. The result is a compact passive isolator capable of very low vertical and horizontal natural frequencies and very high internal structural frequencies. Elastometers are mounted at the structure basis, acting as seismic isolators.

\subsection{Material cleaning and assembling}

An important hazard for the CUORE sensitivity to the $0 \nu \beta \beta$ process, is the background coming from the radioactive contaminations in the cryostat radiation shields, from the mechanical structure of the towers and from the crystals themselves. This background contribution has to be reduced as much as possible. A strict protocol has been adopted for the crystals production, together with a thorough campaign for the selection of the material used to prepare the crystals \cite{crystclean}. This allowed to reduce the presence of environmental radioactivity in the crystals. New cleaning techniques have been developed at the Laboratori Nazionali di Legnaro dell'INFN (LNL) and employed for the copper surfaces, among which: tumbling, electropolishing, chemical etching, magnetron plasma. These have been applied to clean the $10~\mathrm{mK}$ cryostat shield, made out of ferrules and internal tiles, and the parts of the towers. The cleaning techniques are aimed to remove a thin layer of the components' surface, so that dirt and impurities due to handling and manufacturing are removed. Moreover, a strict protocol has been adopted for each step of the CUORE towers construction: temperature sensors gluing, tower assembly, wire bonding, tower storage. All in nitrogen atmosphere and within glove boxes to avoid radioactive recontamination.

\section{Summary}

\begin{figure}[htb]
\centering
\begin{subfigure}{.5\textwidth}
  \centering
  \includegraphics[width=1.\linewidth]{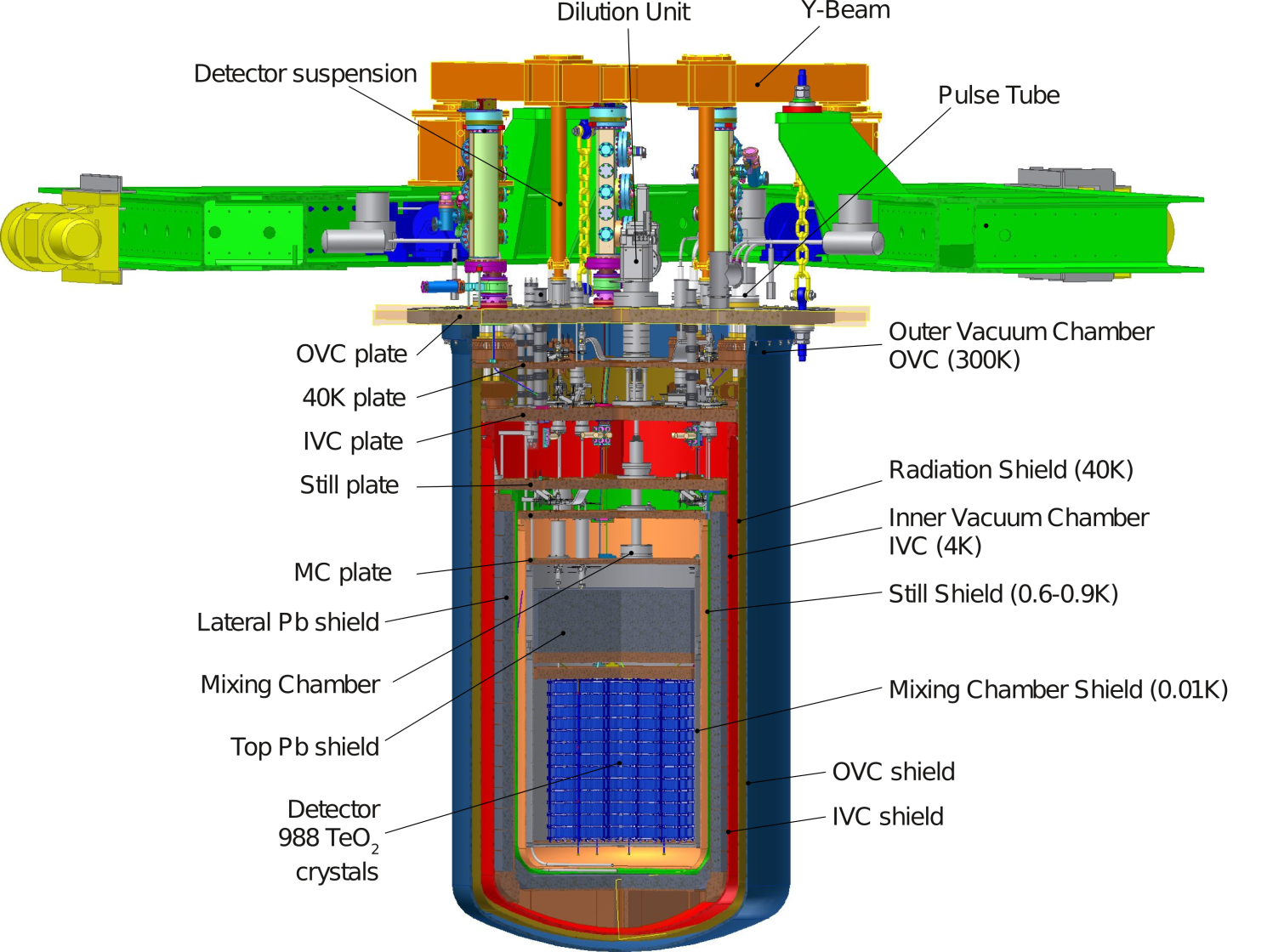}
  \caption{Scheme of CUORE cryostat}
  \label{fig:sub1}
\end{subfigure}%
\begin{subfigure}{.5\textwidth}
  \centering
  \includegraphics[width=1.\linewidth]{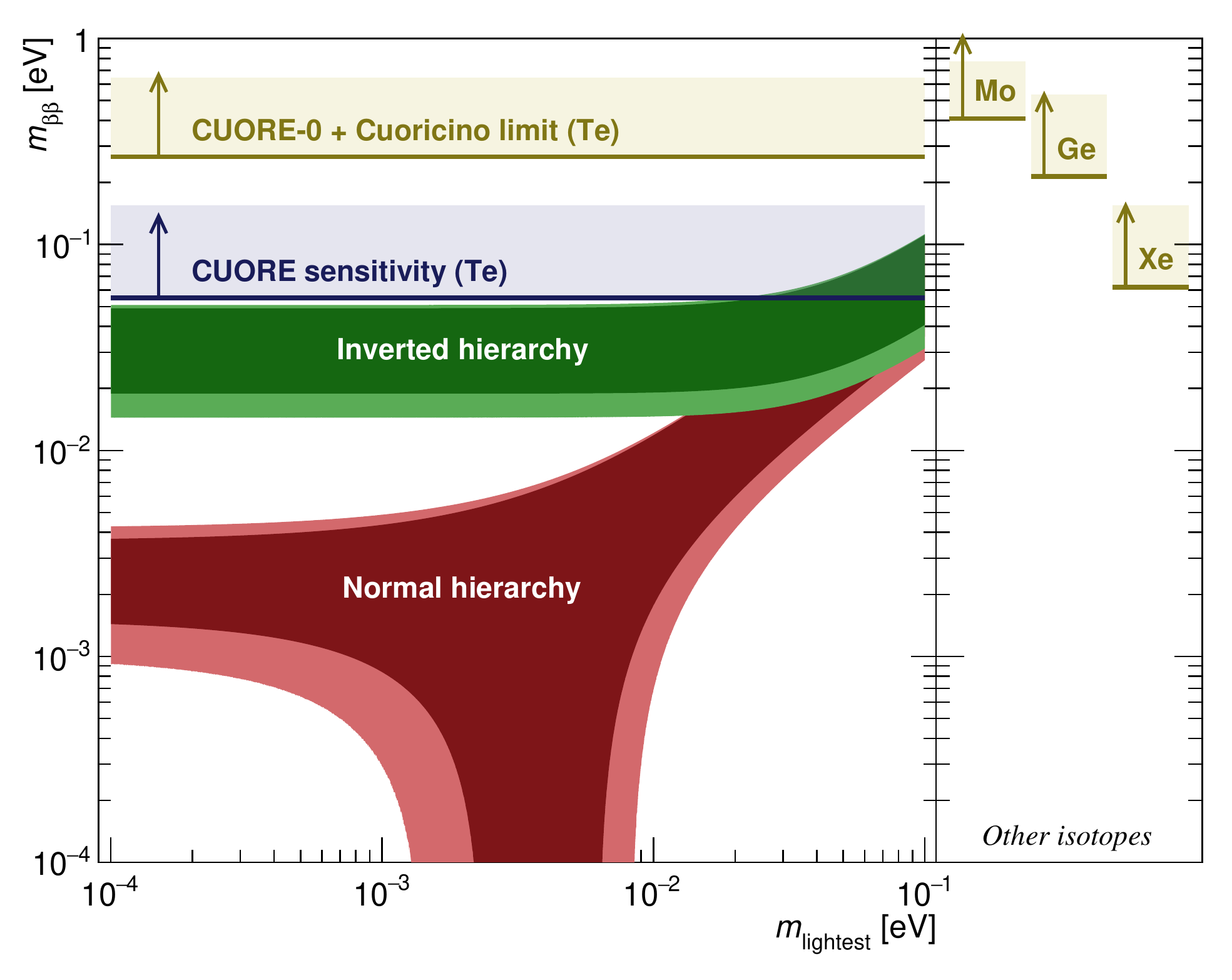}
  \caption{Effective Majorana mass sensitivity}
  \label{fig:sub2}
\end{subfigure}
\caption{}
\label{fig:test}
\end{figure}

A great effort was made to design and develop CUORE in order to achieve, in the ROI, a background of $0.01~\mathrm{counts/(keV\cdot kg\cdot yr)}$ and an energy resolution of $5~\mathrm{keV}$ FWHM. These values will allow to set an upper limit on the $0 \nu \beta \beta$ process half life of $T^{0\nu}_{1/2} > 9.5 \times 10^{25}~\mathrm{yr}$ at $90\%~\mathrm{C.L.}$ in 5 years, interpreted as upper limit on the effective Majorana mass (see fig. \ref{fig:sub2}), and to probe the DAMA/LIBRA Dark Matter positive signal region, exploiting the matched filter technique to extract low energy signals.

The cryostat has been commissioned with full load, but no detector installed, during spring 2016, reaching a stable base temperature of $\sim6.3~\mathrm{mK}$ over more than 70 days. The full detector has been installed on the Tower Support Plate (TSP) during August 2016, and the cryostat closed by the end of November 2016. The detector has been cooled down and commissioning operations have been performed, allowing to debug the full system: check of DAQ and electronic chain, optimization of the noise and mechanical vibrations, optimization of trigger thresholds. The CUORE experiment is now starting to take the first data for physics analyses.



\begin{thebibliography}{99}


\bibitem{0dbd}
Furry W. H., Phys. Rev. {\bf56}, 1184 (1939).
\bibitem{qexp}
Artusa D. R. et al., Adv. High Energy Phys. {\bf2015}, 879871 (2015)
\bibitem{q0exp}
C. Alduino et al., JINST {\bf11}, P07009 (2016).
\bibitem{q0res}
C. Alduino et al., Phys. Rev. Lett. {\bf115}, 102502 (2015).
\bibitem{romanpb}
A. Alessandrello et al., Nucl. Instrum. Meth. B {\bf142}, 163 (1998).
\bibitem{crystclean}
A. Alessandrello et al., J. Cryst. Growth {\bf312}, pp. 2999-3008 (2010).


\end{thebibliography}
\end{document}